\documentclass[fleqn,usenatbib]{mnras}

\usepackage{newtxtext,newtxmath}

\usepackage[T1]{fontenc}
\usepackage{ae,aecompl}

\usepackage{graphicx}	
\usepackage{amsmath}	

\newcommand{\HI}{H\,{\sc i}}
\newcommand{\HII}{H\,{\sc ii}}
\newcommand{\Msun}{~M$_{\odot}$}
\newcommand{\MMsun}{M$_{\odot}$}
\newcommand{\FHI}{F$_{\rm HI}$}

\newcommand{\MHI}{M$_{\rm HI}$}
\newcommand{\Mst}{M$_{\rm *}$}

\newcommand{\kms}{~km\,s$^{-1}$}
\newcommand{\kkms}{km\,s$^{-1}$}
\newcommand{\vsys}{v$_{\rm sys}$}

\newcommand{\vhel}{v$_{\rm hel}$}
\newcommand{\vlg}{v$_{\rm LG}$}
\newcommand{\wfi}{w$_{\rm 50}$}

\newcommand{\MC}{\multicolumn}

\title[Peekaboo galaxy]{Peekaboo: the extremely metal poor dwarf galaxy HIPASS J1131--31}

\author[Karachentsev et al.]{I. D. Karachentsev,$^{1}$\thanks{E-mail: idkarach@gmail.com} 
L.N. Makarova,$^1$
B.S. Koribalski,$^{2,3}$
G.S. Anand,$^{4}$
\newauthor R.B. Tully,$^{5}$
and A.Y.~Kniazev$^{6,7,8}$\\
$^{1}$Special Astrophysical Observatory of the Russian Academy of Sciences, Nizhnij Arkhyz, Karachay-Cherkessia 369167, Russia \\
$^{2}$Australia Telescope National Facility, CSIRO Astronomy and Space Science, P.O. Box 76, Epping, NSW 1710, Australia \\
$^{3}$School of Science, Western Sydney University, Locked Bag 1797, Penrith, NSW 2751, Australia \\
$^{4}$Space Telescope Science Institute, 3700 San Martin Drive, Baltimore, MD 21218, USA \\
$^{5}$Institute for Astronomy, University of Hawaii, 2680 Woodlawn Drive, Honolulu, HI 96822, USA \\
$^{6}$South African Astronomical Observatory, PO Box 9, 7935 Observatory, Cape Town, South Africa \\
$^{7}$Southern African Large Telescope Foundation, PO Box 9, 7935 Observatory, Cape Town, South Africa \\
$^{8}$Sternberg Astronomical Institute, Lomonosov Moscow State University, Universitetskij Pr. 13, Moscow 119992, Russia\\
}

\date{Accepted XXX. Received YYY; in original form ZZZ}

\pubyear{2022}

\begin{document}
\label{firstpage}
\pagerange{\pageref{firstpage}--\pageref{lastpage}}
\maketitle

\begin{abstract}
The dwarf irregular galaxy HIPASS~J1131--31 was discovered as a source of HI emission at low redshift in such close proximity of a bright star that we call it Peekaboo.  The galaxy resolves into stars in images with Hubble Space Telescope, leading to a distance estimate of $6.8 \pm 0.7$~Mpc.  Spectral optical observations with the Southern African Large Telescope reveal HIPASS~J1131--31 to be one of the most extremely metal-poor galaxies known with the gas-phase oxygen abundance 12+log(O/H) = 6.99$\pm$0.16~dex via the direct [OIII] 4363 line method and 6.87$\pm$0.07~dex from the two strong line empirical methods. The red giant branch of the system is tenuous compared with the prominence of the features of young populations in the color-magnitude diagram, inviting speculation that star formation in the galaxy only began in the last few Gyr.  

\end{abstract}




\section{Introduction} 
\label{sec:intro}


\begin{figure*} 
\centering
   \includegraphics[width=17.5cm]{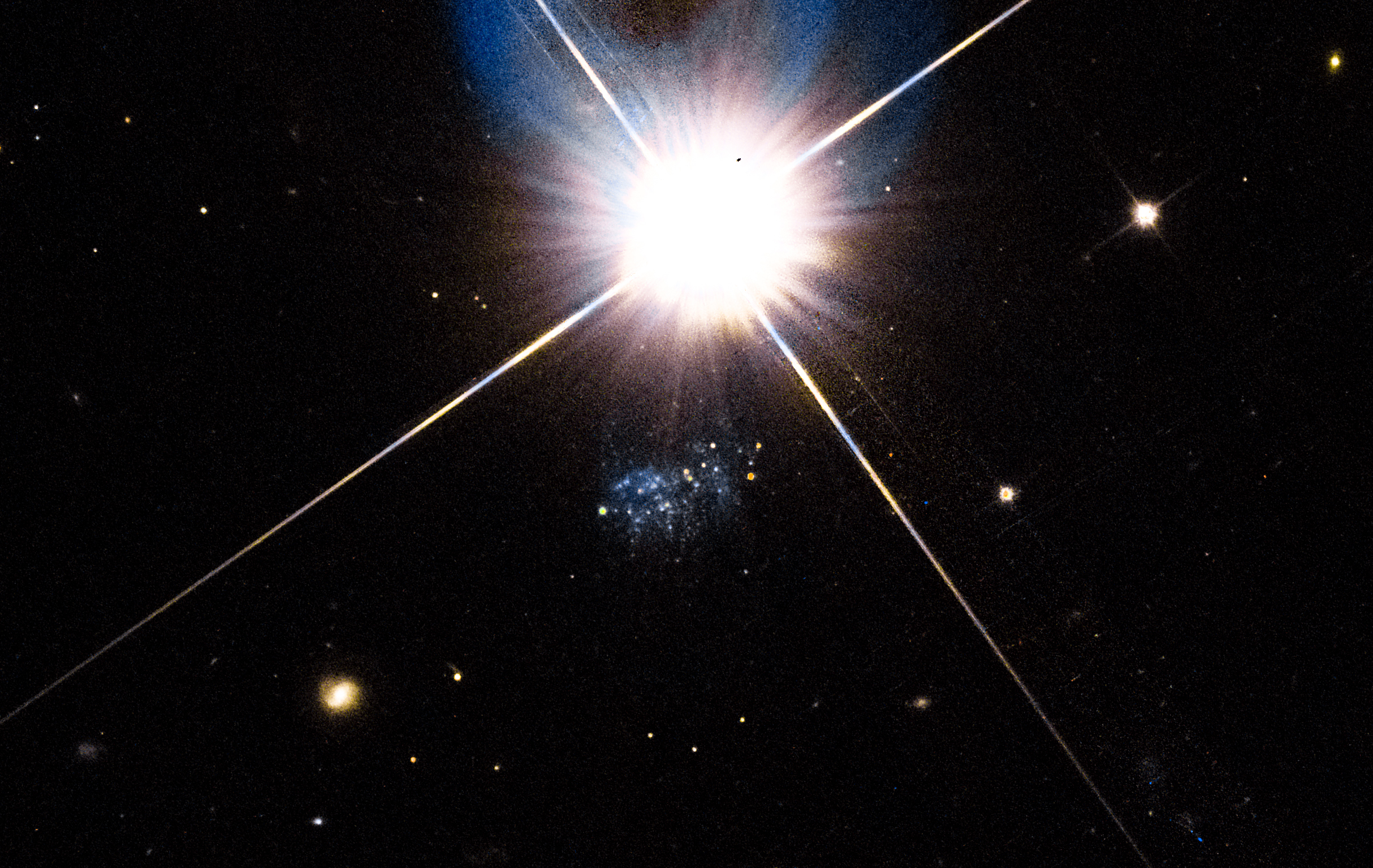}
\caption{Two-colour HST image  of the dwarf galaxy HIPASS~J1131--31, consisting of ACS F606W (blue) and F814W (red). The image dimensions are $\sim$ 70$''$ $\times$ 45$''$, and North is up. The foreground star, TYC~7215-199-1 (10.4 mag), is located $\sim$15 arcsec to the North. Part of its diffraction pattern is included in the displayed image.}
\label{fig:HST-fig1}
\end{figure*}

HIPASS~J1131--31 (PGC\,5060432) is a nearby dwarf irregular galaxy with an \HI\ systemic velocity of 716\kms, originally discovered in the \HI\ Parkes All Sky Survey (HIPASS; \citet{Barnes2001}) by B.S. Koribalski. 
No stellar counterpart could be identified at the time due to the bright foreground star TYC 7215-199-1 (10.4 mag) and its extensive diffraction pattern close to the fitted HIPASS position. Follow-up Australia Telescope Compact Array (ATCA) 21-cm observations, carried out as part of the LVHIS project \citep{Koribalski2018} revealed an unresolved \HI\ source, centered at $\alpha,\delta$(J2000) = $11^{\rm h}\,31^{\rm m}\,34.6^{\rm s}$, --31\degr\,40\arcmin\,28.3\arcsec. \citet{Koribalski2018} found this position to coincide with a compact blue dwarf galaxy in GALEX far-ultraviolet (FUV) images. 

HIPASS\,J1131--31 was subsequently imaged as part of a Hubble Space Telescope (HST) SNAP program given the possibility of proximity based on the observed 21cm redshift.  Sure enough, a dwarf galaxy that resolves well into stars emerged from the glare of the bright foreground star.  Analysis of the color-magnitude diagram of resolved stars and spectroscopic follow up reveals the galaxy to have the extreme properties of a young system. We call this dwarf the Peekaboo galaxy because of the way it has been hiding and because of its potential importance.

The closest neighbours of HIPASS~J1131--31 are the dwarf irregular galaxy HIPASS~J1132--32 (\vsys\ = 699\kms) and 
the large spiral galaxy NGC~3621 (HIPASS~J1118--32; 730\kms) 
at projected distances of 80 and 182 arcmin, respectively \citep{Koribalski2004,Meyer2004}. ATCA \HI\ images of both galaxies were also obtained as part of the LVHIS project \citep{Koribalski2018}. The distances of NGC~3621 ($6.70 \pm 0.37$ Mpc) and HIPASS~J1132--32 ($5.47 \pm 0.15$ Mpc) were obtained by \citet{Ferrarese2000} via a supernova luminosity and \citet{Anand2021} via a tip of the red giant branch magnitude, respectively. Based on the likely association of HIPASS~J1131--31 with NGC~3621 we initially assigned the same distance, i.e. $D_{\rm mem}$ = 6.7 Mpc \citep{Koribalski2018}. Using ATCA we measured \FHI\ = 1.13 Jy\kms,  corresponding to \MHI\ = $1.2 \times 10^7$\Msun. By applying the \HI\ mass-size relation \citep{Wang2016} we infer an approximate \HI\ diameter of $\sim$1.7 kpc ($52''$).

In this paper we analyse new optical and \HI\ images of the dwarf galaxy HIPASS~J1131--31 = Peekaboo. Brief summaries of our HST optical, ATCA \HI\ observations and SALT spectroscopy are given in Section~2, followed by our analysis of the properties of Peekaboo in Section~3. The galaxy environment is discussed in Section~4. Properties of other similar compact gas-rich dwarf galaxies in the Local Volume are described in Section~5, and our conclusions are given in Section~6.

\section{Observations and Data Analysis}
\label{sec:obs}

\subsection{HST}

The dwarf galaxy HIPASS~J1131--31 was observed with the Hubble Space Telescope (HST) Advanced Camera for Surveys (ACS) on 14 July 2020 for 760 seconds in each of the $I$- and $V$-bands as a part of the "Every Known Nearby Galaxy" survey (SNAP-15922, PI R.B. Tully). The respective central wavelengths were 5921.89\AA\ (F606W) and 8045.54\AA\ (F814W). The pixel size is 0.05 arcsec. Figure~1 shows the HST two-colour composite image. The galaxy lies 15 arcsec south of a bright star and is obscured by the star's halo in Palomar Observatory Sky Survey images.
The area of the galaxy is compromised in the HST images because the bright star is not excluded from the ACS field in the SNAP positioning.  Nonetheless, the brightest stars are usefully resolved.

\begin{figure} 
\centering
\includegraphics[width=9cm]{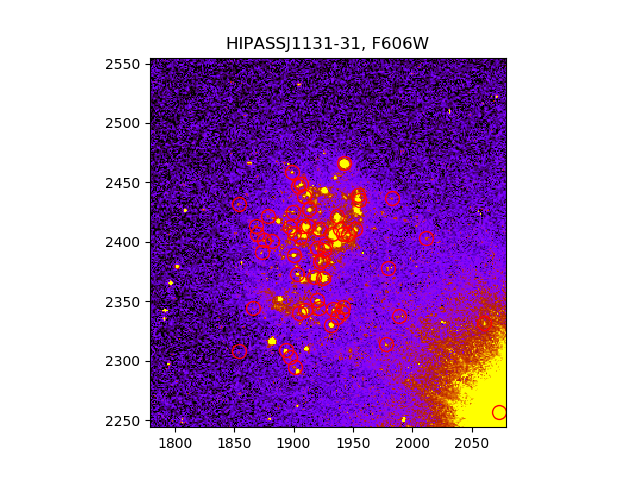}
\includegraphics[width=7cm]{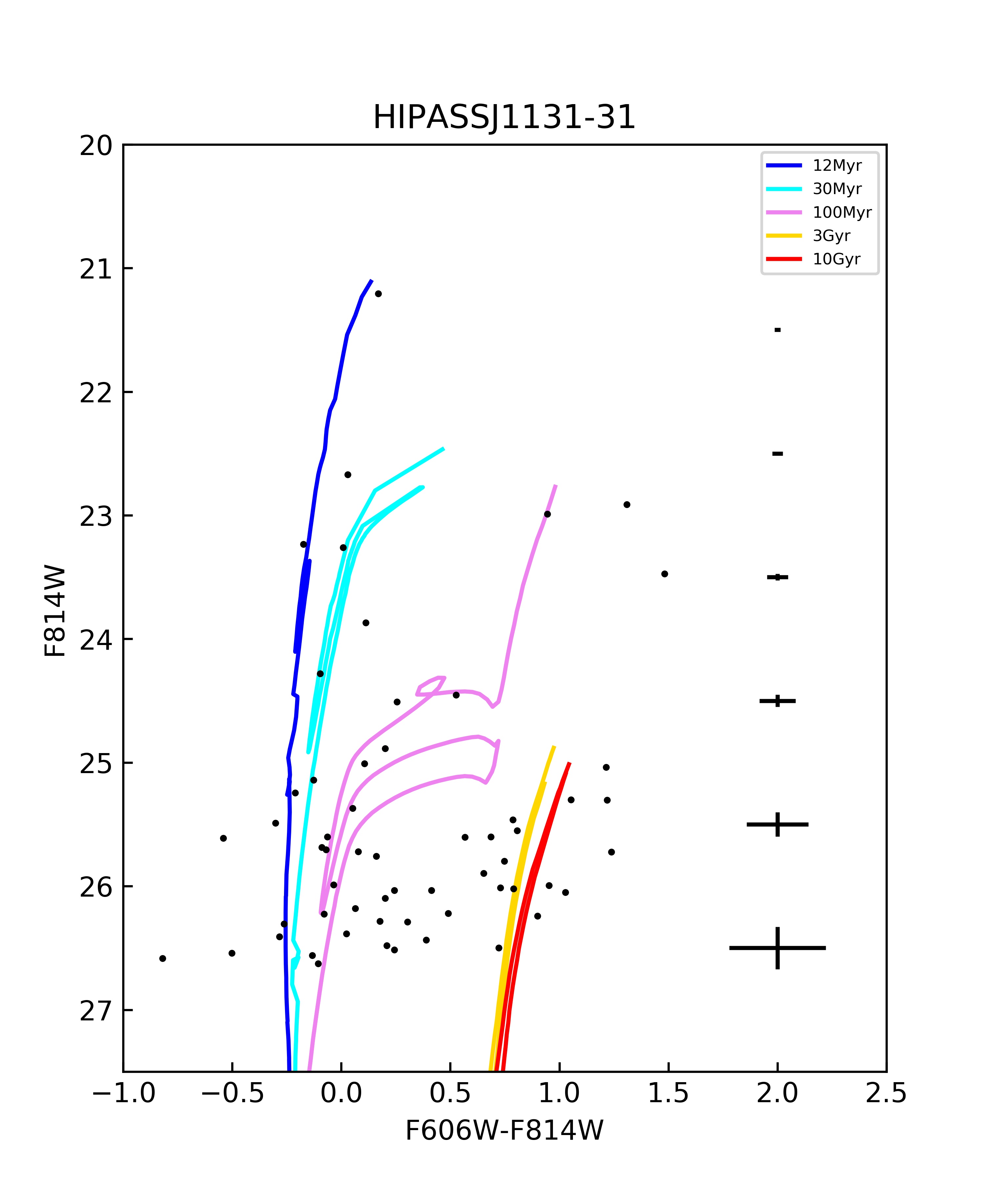}
\caption{{\bf Top:} HST/ACS F606W image of the Peekaboo galaxy with the indicated pixel coordinates. North is in the direction toward the bottom-right corner. Stars detected within the galaxy are marked with open red circles. ---
{\bf Bottom:} HST/ACS derived colour magnitude diagram of the Peekaboo galaxy. We overlay PARSEC stellar isochrones \citep{Bressan2012,Marigo2013} assuming a metallicity of $Z = 0.0002$ and a distance of 7.0 Mpc.}
\label{fig:cmd}
\end{figure}

We used the ACS module of the DOLPHOT package by \citet{Dolphin2000, Dolphin2016} to perform photometry of resolved stars based on recommended parameters. Only stars with a signal-to-noise ratio at least four in both filters were included in the analysis. For further analysis, we limited ourselves to the consideration of stars located within the immediate vicinity of the galaxy, in a square with dimensions 300 by 300 pixels or 15 by 15 arcsec (see Fig.~\ref{fig:cmd}, top panel). The measured stars are indicated with open circles, and the resulting colour-magnitude diagram for them in F606W--F814W versus F814W is shown in the bottom panel. Despite the strong crowding of the stellar population, 58 stars are well-recovered. As can be seen, the galaxy is dominated by blue main sequence and blue loop stars. There are remarkably few old red stars given that the HST imaging resolves stars $\sim 1.5$ mag fainter than the putative tip of the red giant branch. We overlay PARSEC stellar isochrones for stellar ages of 12, 30, 100, 500 Myr, and 10 Gyr assuming a metallicity of $Z = 0.0002$ (see Section~\ref{txt:SALT_spec})
and a distance of 7.0 Mpc. The brightest blue star on the diagram corresponds to the age of 12 Myr and the mass of 17\Msun.

The red giant branch is remarkably sparse.  The isochrone fit seen in Fig.~\ref{fig:cmd} assumes that the brightest red stars lie near the tip of the red giant branch (TRGB).  However with a normal old population there would be an abundance of fainter red giant stars.  
We adopt the magnitude of the TRGB to be $25.2 \pm 0.2$ mag. Following the zero-point calibration of the absolute magnitude of the TRGB developed by \citet{Rizzi2007} and the rescaling that brings TRGB distance moduli 0.05 mag closer by \citet{Tully2022}, we obtained M(TRGB) = --4.06. Assuming the colour-excess $E(B-V)$ = 0.062 from \citet{Schlafly2011} as foreground reddening, we derive the true distance modulus of $29.2 \pm 0.2$ or the distance $D = 6.8 \pm 0.7$ Mpc.  Given the low population of the RGB, the brightest red giant stars may fall short of the tip of the RGB whence our distance estimate should be viewed as an upper limit. It is seen that the other six galaxies in a loose association with Peekaboo galaxy identified in Table~\ref{tab:ngc3621-group-tab1} all with well determined TRGB distances, have the averaged distance of $6.2\pm0.7$~Mpc.

\begin{figure*} 
\centering
   \includegraphics[width=12cm]{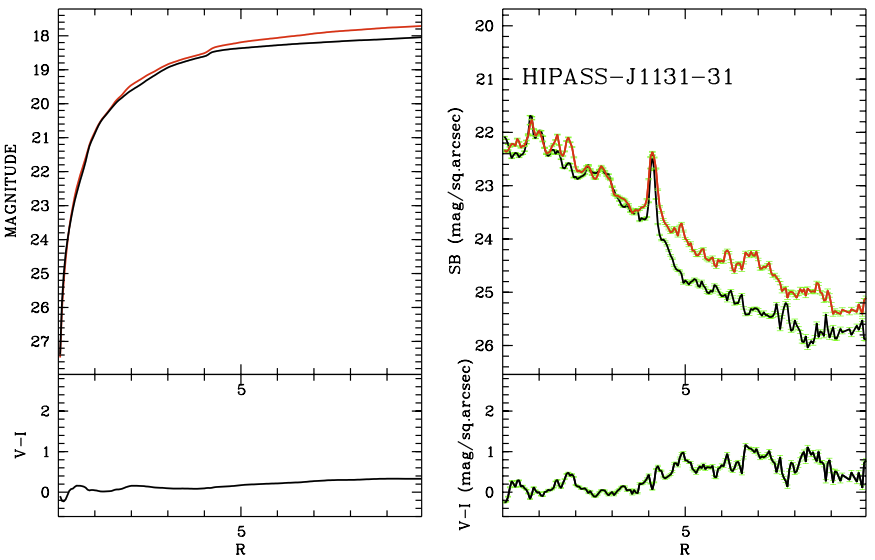}
\caption{Surface brightness photometry of the galaxy HIPASS~J1131--31 as obtained from the HST ACS $V$- and $I$-band images. --- {\bf Left:} The integral magnitude of the galaxy in the $I$ and $V$ filters and the integral colour index as a function of the circular radius $R$ in arcseconds. {\bf Right:} The measured profile of the galaxy brightness in the $I$ and $V$ bands and variations of the colour index ($V - I$) with dependence on $R$ in arcseconds. }
\label{fig:surf}
\end{figure*}

The images of Peekaboo obtained with the HST/ACS were used to perform the surface photometry of the galaxy following the method by \citet{Sharina_Kryzhanovsky2022}. Figure~\ref{fig:surf} shows the results. The upper left panel presents behavior of the integral magnitude of the galaxy in the $I$ and $V$ filters depending on the circular radius $R$ in arcseconds. The bottom left panel reflects the variations with the radius of the integral colour index. From these data, the integral magnitudes of the galaxy within $R = 10''$ are $V$ = 18.07 and $I$ = 17.73 mag.  Taking into account the relation $B - V$ = 0.85 ($V - I$) -- 0.20  \citep{Makarova_Karachentsev1998}, we obtain $B$ = 18.16 mag. The right panels of Fig.~3 demonstrate the measured profile of the brightness of Peekaboo in $I$ and $V$ bands as well as the variations of the colour index ($V-I$). 

Peekaboo galaxy is clearly detected in GALEX FUV images, but not detected in the NUV images. It is twice catalogued as GALEX J113134.7--314026 with a FWFM of 10.8 arcsec, and GALEX J113134.5--314025 with a FWHM of 9.7 arcsec.
 
\subsection{ATCA}
\begin{figure*} 
\centering
   \includegraphics[width=12cm]{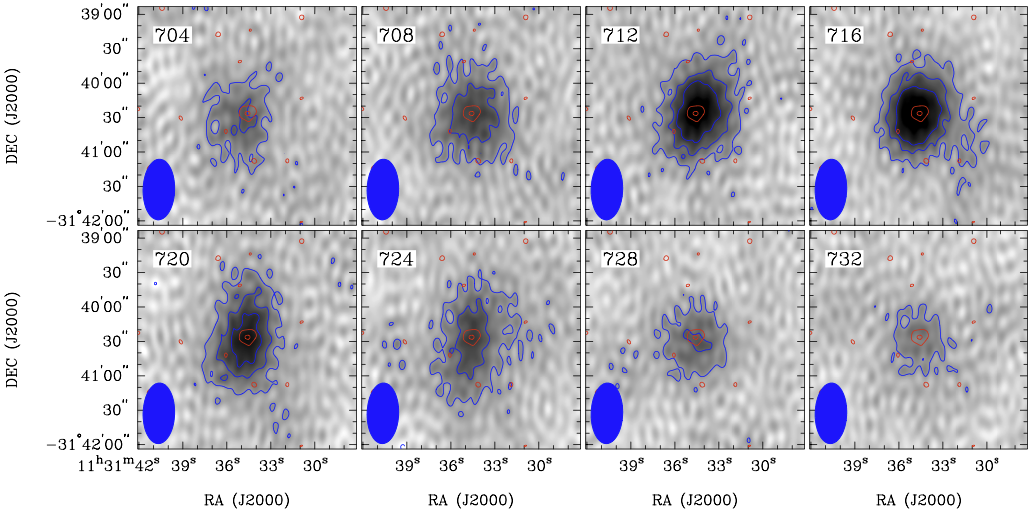}
\caption{ATCA \HI\ channel maps of the dwarf galaxy HIPASS~J1131--31 using $r = 0.5$ weighting (beam = $52.7'' \times 27.4''$). The \HI\ contour levels are --6, 6, 12 and 18 mJy\,beam$^{-1}$ (coloured blue); the heliocentric \HI\ velocity is shown in the top left corner of each panel. For reference we overlaid the GALEX FUV emission in red contours.}
\label{fig:ATCA-HIchannels05}
\end{figure*}

\begin{figure*}
\centering
   \includegraphics[width=12cm]{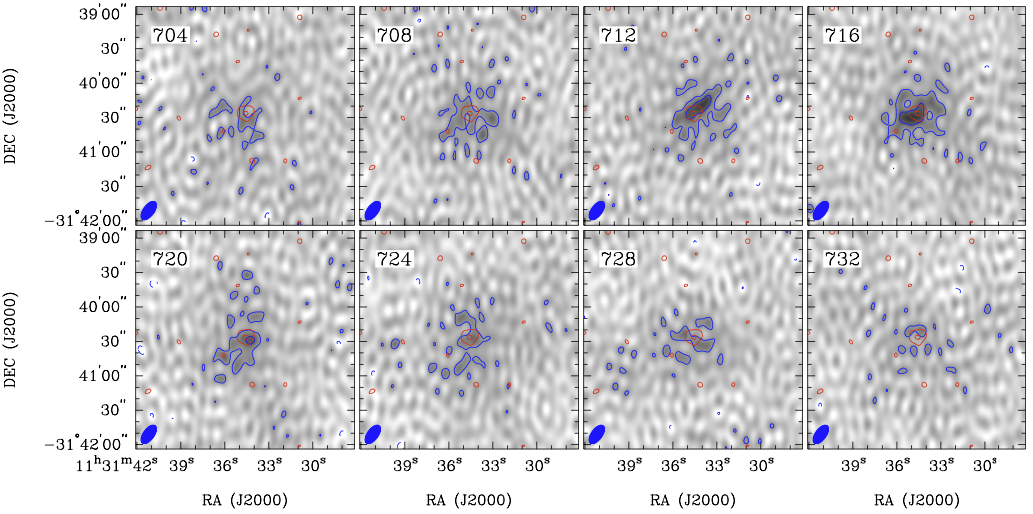}
i\caption{ATCA \HI\ channel maps of the dwarf galaxy HIPASS~J1131--31 using $r = 0$ weighting (beam = $18.6'' \times 9.3''$, $PA$ = --36.1 degr). The \HI\ contour levels are --6, 6, 12 and 18 mJy\,beam$^{-1}$ (coloured blue); the heliocentric \HI\ velocity is shown in the top left corner of each panel. For reference we overlaid the GALEX FUV emission in red contours.}
\label{fig:ATCA-HIchannels0}
\end{figure*}

We use the same Australia Telescope Compact Array (ATCA) data of HIPASS~J1131--31 = Peekaboo analysed by \citet{Koribalski2018} as part of the Local Volume \HI\ Survey (LVHIS). These consists of three configurations (EW353, 750C and 1.5B), each observed for $\sim$12h. Here we analyse new high-resolution \HI\ maps, marginally resolving Peekaboo. Using robust weighting we obtain an angular resolution of $18.6'' \times 9.3''$ (for $r = 0$) and $52.7'' \times 27.4''$ (for $r = 0.5$). The measured rms noise is $\sim$2 mJy\,beam$^{-1}$. The derived \HI\ channel maps of the galaxy with different resolution are presented in Figs.~\ref{fig:ATCA-HIchannels05} \& \ref{fig:ATCA-HIchannels0}. 

Fitting a Gaussian to the resolved \HI\ intensity distribution of Peekaboo gives a deconvolved size of $\sim$24\arcsec\ $\times\ 12\arcsec$ with a position angle of $PA \sim 120\degr$. While highly uncertain, this suggests the \HI\ size is approximately twice the optical size.

\subsection{SALT spectroscopy}
\label{txt:SALT_spec}

\begin{figure*}
\centering
   \includegraphics[width=12.9cm]{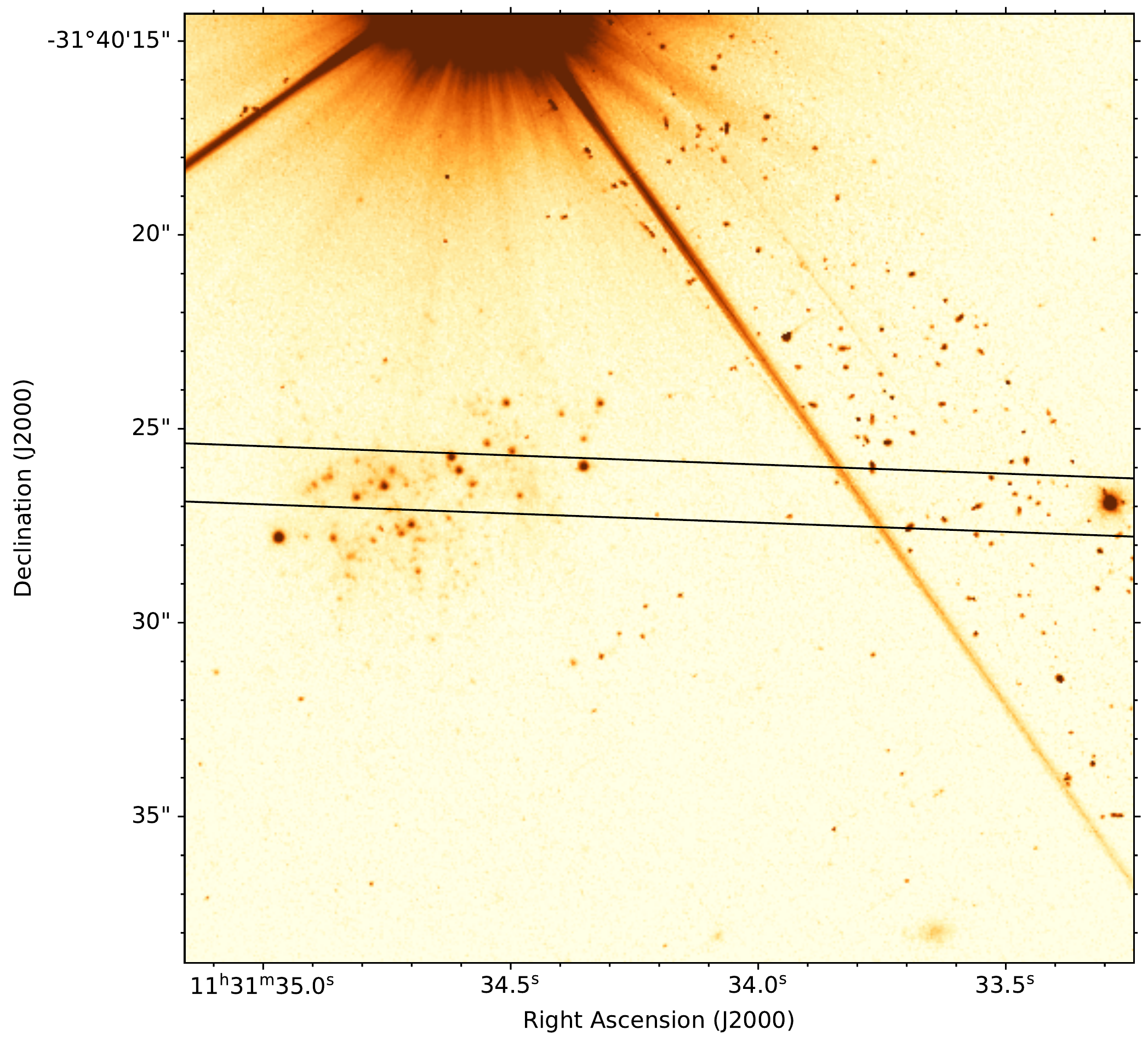}
   \includegraphics[width=12.9cm]{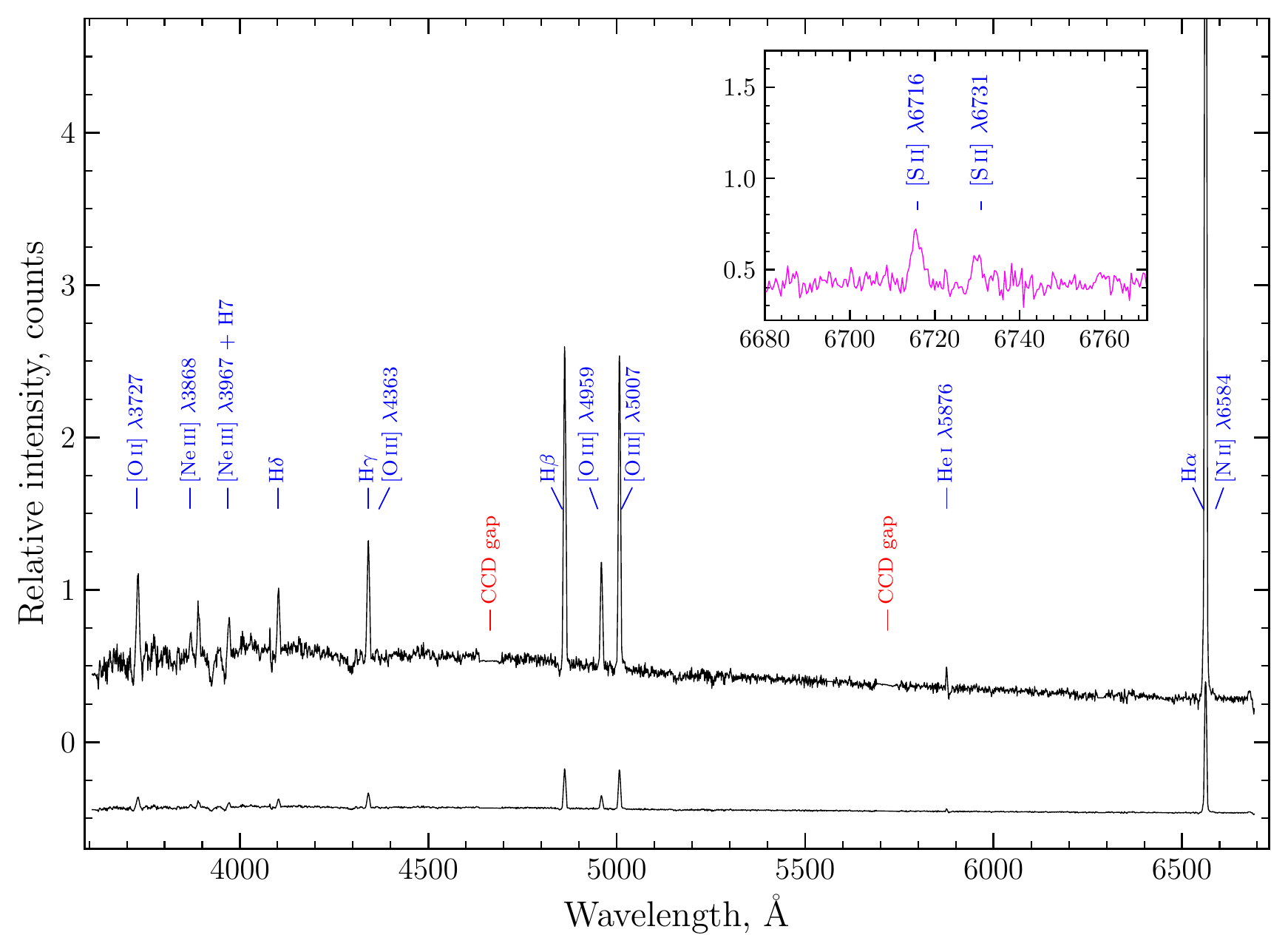}
\caption{{\bf Top:} Long-slit position for the spectral observations of HIPASS~J1131--31. The slit width is 1\farcs5. The relatively bright star that is located in the slit was used as the reference star to be sure that specific part of the galaxy will be observed.
{\bf Bottom:} SALT optical spectrum of  HIPASS~J1131--31 with the most prominent lines indicated. The spectrum at the bottom is scaled by 1/8 and shifted to show the relative intensities of the strong lines.}
\label{fig:SALT_spec}
\end{figure*}

We carried out long-slit optical spectroscopic observations of Peekaboo galaxy using the Robert Stobie Spectrograph \citep[RSS;][]{Burgh03,Kobul03} installed at the Southern African Large Telescope \citep[SALT;][]{Buck06,Dono06} on 22 February, 2022. The VPH gratings PG0900 and PG1800 were used with the long slit of 1.5\arcsec\ by 8\arcmin\ to cover the full range from 3600\AA\ to 6800\AA\ with the resulting spectral resolution of FWHM$\sim$6.0\AA\ for most of the lines observed with grating PG0900 and FWHM$\sim3.0$\AA\ for the lines [SII] $\lambda\lambda~6717,6731$\AA\ observed with grating PG1800. All spectral data were obtained with a binning factor of 2 for both spatial and spectral directions to give a final spatial sampling of 0\farcs25~pixel$^{-1}$ and spectral sampling of 0.97\AA\,pixel$^{-1}$ for PG0900 and 0.40\AA\,pixel$^{-1}$ for grating PG1800. The position angle ($PA$) for observations was selected to be 87.88\degr\ as shown in the top panel of Figure~\ref{fig:SALT_spec}. Two 1300\,s exposures for each configuration were obtained under seeing conditions of 1.5 arcsec. 

Since SALT is equipped with an Atmospheric Dispersion Compensator (ADC), there is no effect due to atmospheric dispersion at an arbitrary long-slit $PA$. Ar and Ne lamp spectra were taken immediately after science frames for the wavelength calibration. Spectrophotometric standards were observed during twilight as a part of the SALT standard calibrations plan. As a consequence of the SALT design, the pupil of the telescope moves during the track and exposures, thereby constantly changing the effective area of the telescope.  Because of this, accurate absolute photometry and spectrophotometry are not feasible, but the relative flux calibration could be used. Hence, the relative distribution of energy in spectra could be obtained with SALT data.

The primary data reduction was done with the SALT science pipeline \citep{Cra2010} which includes bias and overscan subtraction and gain correction for each CCD amplifier, cross-talk correction and finally mosaicing. The long-slit reduction was done with the RSS pipeline described by \citet{Kniazev2022}. The 1D spectra extraction, emission lines measurement and calculation of chemical abundances was done in the way described by \citet{Kniazevetal2008}. The fully reduced 1D spectrum of Peekaboo is shown in the bottom panel
of Figure~\ref{fig:SALT_spec} with labeling of the most prominent lines.
The insert in the figure shows a portion of the spectrum taken with PG1800 for the spectral region around the lines [SII] $\lambda\lambda~6717,6731$\AA.  Since the RSS detector is a mosaic and consists of three CCDs, the positions of two gaps between these CCDs are also shown.

Errors in lines intensities have been propagated in the calculations 
of the reddening correction, electron temperatures and densities and have been propagated
to the elemental abundance errors.
The observed emission line intensities $\rm F(\lambda)$ relatively to the F(H$\beta$)
line and the intensities $\rm I(\lambda)$ corrected
for interstellar extinction and underlying stellar absorption
are presented in Table~\ref{t:Intens}.
The equivalent width $EW$(H$\beta$) of the H$\beta$ line,
the calculated absorption equivalent widths $EW$(abs) of the Balmer lines,
the extinction coefficient $C$(H$\beta$)
(a sum of the internal extinction in Peekaboo and foreground extinction in the Milky Way) and the subsequently recalculated $E(B-V)$ value are also listed there.

The object oxygen abundance, 12 + log(O/H) dex, was determined in three ways.  The first way is the classic $T_{\rm e}$ method, which is based on the direct calculation
of $T_{\rm e}$(OIII) making use of the intensity of the faint auroral line [\ion{O}{iii}]~$\lambda4363$
\citep[see the full description of the method used in][]{Kniazevetal2004,Kniazevetal2008}.
The measured ratio [\ion{S}{ii}]~($\lambda$6716/$\lambda$6731)
$\simeq2$ implies a very low number density of ionised gas of $\le$10~cm$^{-3}$.
The calculation of density $N_{\rm e}$(SII) is not possible for a value of
\mbox{[\ion{S}{ii}]~($\lambda$6716/$\lambda$6731)$\,\ge1.4$}, so
we used in our calculations the value $N_{\rm e}$(SII) = 10$\pm$10~cm$^{-3}$
as an upper limit of the density of ionised gas in Peekaboo galaxy.
Since density and its error fold into iterations of the $T_{\rm e}$(OIII) calculation
\citep[see equation (3) in][]{Kniazevetal2008}, and also the accuracy of the [\ion{O}{iii}]~$\lambda4363$
measurement is poor (only 2.5 sigma), the final oxygen abundance accuracy measured in the classic method is low.
We label $T_{\rm e}$(OIII) estimated by this way as $T_{\rm e}$(OIII)$_{4363}$.
This direct method gives the abundance 12 +log(O/H)$_{4363}$ dex = 6.99$\pm$0.16~dex.

Secondly, we used  a so-called semi-empirical method \citep{Izotov2007}.
This method at the beginning calculates $T_{\rm e}$(OIII) with use of an empirical
calibration (for example, a correlation between $T_{\rm e}$ and a parameter $R_{\rm 23}$) and then
uses this calculated $T_{\rm e}$ and its errors in a procedure similar to the classical direct
method as described above. In our case for the estimation of $T_{\rm e}$(OIII) we used a
ratio from \citet{Izotov2007} which was modified by \citet{Pustilniketal21} to take
into account the variance of the excitation parameter $Q_{\rm 23}$.
We identify $T_{\rm e}$(OIII) estimated by this way as $T_{\rm e}$(OIII)$_{P21}$.
In the case of this semi-empirical method the calculated error of $T_{\rm e}\
$(OIII)$_{P21}$
is much smaller since the estimation of $T_{\rm e}$ is based on ratios of strong emission lines.
Here, the number density $N_{\rm e}$(SII) is not included in the calculation of
$T_{\rm e}$(OIII)$_{P21}$. Hence, in case of the Peekaboo spectrum,
the calculated error of the oxygen abundance with this semi-empirical method is considerably smaller
compare to the classic $T_{\rm e}$ method.
In this way, we obtained the oxygen abundance of 12 +log(O/H)$_{P21}$ = 6.80$\pm$0.05~dex.

As the third way to estimate the oxygen abundance of HIPASS~J1131--31,
we use the variant of an empirical 'strong-lines' method suggested by
\citet{Izotov2019} and \citet{IzThGu2021}. This empirical method was developed
specifically for the range of the lowest values of O/H (12 + log(O/H) $\le$ 7.5~dex).
With use of this method we obtained the oxygen abundance
of 12 +log(O/H)$_{ITG21}$ = 6.93$\pm$0.02~dex.
Since this empirical method is based on ratios of strong emission lines only,
the final statistical error is even less than for two previous calculations.  However, we are here in a regime of unexplored systematic uncertainties.


All temperatures and abundances calculated for Peekaboo with different methods
as well as their estimated errors are presented in Table~\ref{t:Intens}.
The three estimates of the oxygen abundance are in satisfactory statistical agreement. As
a result, we get two values for the oxygen abundance: (1) the value 12 +log(O/H) = 6.99$\pm$0.16~dex with the direct method, and (2) 6.87$\pm$0.07~dex as an average of the two strong line empirical methods, with smaller statistical errors but acknowledged unknown possibilities of systematic errors.

The oxygen abundance obtained for Peekaboo is at the low edge of values for the class of objects called Extremely Metal-Poor (XMP) galaxies. Our
calculated abundance can be compared with those for other known XMP dwarfs: 7.17$\pm$0.04 for I~Zw~18
\citep{Skillman2013}, 7.06$\pm$0.03 for the Leoncino Dwarf \citep{Aver2022}, 6.95$\pm$0.06 for
AGC~124629 and 7.05$\pm$0.05 for AGC~411446 \citep{Pustilnik2020}, and 6.98$\pm$0.02 for J0811+4730
\citep{Izotov2018}. Therefore, Peekaboo turns out to be one of the most metal-poor
star-forming dwarf galaxies known so far. With the absolute magnitude of $M_B = -11.3$ (see Table 5 below), Peekaboo deviates by 0.5~dex from the luminosity - metallicity relation for nearby dwarf star-forming galaxies shown in Fig.~8 of \citet{McQuinn2020} and Fig.~3 in \citet{Pustilnik2020}. Its offset is less than those for I~Zw~18 and DDO~68, but more than for Leoncino.
Of the well resolved XMP objects with weak RGB features, Peekaboo galaxy is the
nearest, being about twice as close as the Leoncino dwarf or DDO~68. The interesting question here is whether
there is any underlying old population or whether Peekaboo is an entirely young galaxy. 

 
\begin{table}
    \centering{
\caption{Peekaboo galaxy line intensities}
\label{t:Intens}
\begin{tabular}{lcc} 
\hline
            \rule{0pt}{10pt}
            $\lambda_{0}$(\AA) Ion   & F($\lambda$)/F(H$\beta$)&I($\lambda$)/I(H$\beta$) \\ \hline
            3727\ [O\ {\sc ii}]\     & 0.329$\pm$0.042 & 0.352$\pm$0.049 \\
            3868\ [Ne\ {\sc iii}]\   & 0.050$\pm$0.019 & 0.053$\pm$0.022 \\
            4101\ H$\delta$\         & 0.175$\pm$0.018 & 0.269$\pm$0.034 \\
            4340\ H$\gamma$\         & 0.398$\pm$0.022 & 0.473$\pm$0.031 \\
            4363\ [O\ {\sc iii}]\    & 0.025$\pm$0.010 & 0.024$\pm$0.011 \\            
            4861\ H$\beta$\          & 1.000$\pm$0.017 & 1.000$\pm$0.021 \\
            4959\ [O\ {\sc iii}]\    & 0.325$\pm$0.017 & 0.299$\pm$0.017 \\
            5007\ [O\ {\sc iii}]\    & 0.990$\pm$0.034 & 0.906$\pm$0.033 \\
            6563\ H$\alpha$\         & 3.353$\pm$0.087 & 2.709$\pm$0.082 \\
            6584\ [N\ {\sc ii}]\     & 0.029$\pm$0.005 & 0.023$\pm$0.005 \\
            6716\ [S\ {\sc ii}]\     & 0.043$\pm$0.002 & 0.034$\pm$0.002 \\
            6731\ [S\ {\sc ii}]\     & 0.021$\pm$0.001 & 0.017$\pm$0.001 \\
            & & \\
            C(H$\beta$)\ dex          & \MC {2}{c}{0.20$\pm$0.03} \\
            $E(B-V)$\ mag             & \MC {2}{c}{0.14$\pm$0.02} \\
            EW(abs)\ \AA\             & \MC {2}{c}{2.25$\pm$0.33} \\   
            EW(H$\beta$)\ \AA\        & \MC {2}{c}{  30$\pm$ 2}   \\ 
            \hline
            6716/6731\ [S\ {\sc ii}]\ & \MC {2}{c}{2.00$\pm$0.25} \\
            $N_e$(SII)(cm$^{-3}$      & \MC {2}{c}{10$\pm$10}       \\
            \hline
            %
            $T_{\rm e}$(OIII)$_{4363}$(K)\  &\MC{2}{c}{17,555$\pm$4120} \\
            $T_{\rm e}$(OIII)$_{P21}$(K)\   &\MC{2}{c}{23,000$\pm$1760} \\
            12+$\log$(O/H)$_{4363}$\ dex    & \MC {2}{c}{6.99$\pm$0.16} \\
            12+$\log$(O/H)$_{P21}$\ dex     & \MC {2}{c}{6.80$\pm$0.05} \\
            12+$\log$(O/H)$_{ITG21}$\ dex   & \MC {2}{c}{6.93$\pm$0.02} \\
\hline
\end{tabular}
    }
\end{table}

\section{Global Properties}

Basic optical and \HI\ properties of Peekaboo galaxy are presented in Table~\ref{tab:galaxy-prop}. They include the galaxy centre position, heliocentric velocity, \HI\ line width and flux, integrated $V$- magnitude and ($V-I$) colour, effective (half-light) radius (averaged via $V$ and $I$ frames), the central surface brightness in $V$-band, optical axis ratio and position angle, as well as the distance-dependent parameters of the galaxy: absolute $V$- magnitude, effective radius in pc, stellar and hydrogen mass, integral and specific star formation rates.

 We estimated the stellar mass of Peekaboo, \Mst, via its integral $I$-band luminosity and the colour index (V-I). We used the $I$-band magnitude $I= 17.61$ which is inferred from the HST photometry after the correction 
of $A_I= 0.12$ mag for Galactic extinction. With $M_I = -11.53$ and $I_{\odot} = 4.10$ (Vega system) the integral luminosity of the 
galaxy is $L_I= 1.78\times 10^6 L_{\odot}$. In Fig.3 of \citet{Kourkchi2022}, values of $M_*/L_I$ for late-type galaxies are presented 
as a function of their ($g - i$) color: $log(M_*/L_I) = 0.44 (g - i) -0.23 \pm0.07$ with a typical rms scatter on this relation of 0.07 dex. Taking for Peekaboo the integrated colour index $(g - i)_0 = 0.2\pm0.1$ corrected for extinction, we get $log(M_*/L_I) = -0.15\pm0.08$ or
$M_*/L_I = 0.71\pm0.13$. This gives the galaxy stellar mass estimate of $(M_*/M_{\odot}) =1.26\times 10^6$.
In principle, the mass-to-$I$-band-light ratio may be improved based on the galaxy SED fitting.
However, the estimate of the stellar mass via SED must inevitably be uncertain for such a dwarf with luminosity dominated by young stars and poor information regarding an underlying population that is expected to dominate the mass.

The hydrogen mass estimate is \MHI\ = $2.356~ \times 10^5 D^2$ \FHI. The integral star formation rate (SFR) is calculated from the galaxy FUV- magnitude with a correction for Galactic extinction $A_B$ \citep{Schlafly2011} of log$(SFR)= 2.78 - 0.4(FUV - 1.93 A_B) + 2 {\rm log}(D)$, where $D$ is the galaxy distance expressed in Mpc.

As can be seen from these data, Peekaboo is a gas-rich dIrr galaxy, whose hydrogen envelope 
extends over 2 optical (and also FUV) dimensions of its body. 
Taking helium into account with multiplier 1.33, the ratio of galaxy mass of gas to its stellar mass is $M_{gas}/M_* = 13$. If plotted in Fig.6 of the \citet{Kourkchi2022} paper, Peekaboo galaxy would lie at the extreme lower left of the log$(M_*/M_{gas})$~ vs.~log$(W_{50})$ plot.

Within a cocoon of gas, the galaxy exhibits high star-forming activity. With the observed SFR, the galaxy is able to reproduce its current stellar mass over only 1.1 Gyr. It has enough gas reserves to support the observed SFR during another 18 Gyr. Peekaboo may be a case of a young galaxy at an early stage of its cosmic evolution. 
Relative to the nearest massive spiral galaxy NGC~3621, Peekaboo is located at a projected separation of 370 kpc, which is larger than the virial radius of NGC~3621 of $\sim$190 kpc. Consequently, the high star-formation activity of the dwarf galaxy is an intrinsic property, independent of the tidal influence of its massive neighbor.

\begin{table}
\centering
\caption{Some properties of the dwarf galaxy HIPASS~J1131--31}
\begin{tabular}{lcc}
\hline
  source name        & HIPASS J1131--31/Peekaboo    &    Ref. \\
\hline
  alternate names    & GALEX J113134.7--314026 \\ 
                     & GALEX J113134.5--314025 \\ 
\hline

\HI\ centre position & $11^{\rm h}\,31^{\rm m}\,34.6^{\rm s}$ & [1]\\ 
  & $-31^{\circ}\,40'\,28.3''$ \\ 
optical type         & dIrr        & [1] \\
\vhel\ [\kkms]       & $716 \pm 1$ & [1] \\ 
\vlg\  [\kkms]       & 433 & \\
\wfi\ [\kkms]        & 29  & [here] \\ 
\FHI\ [Jy\kms]       & 1.13 & [1] \\
approx. \HI\ size    & $\sim$24\arcsec\ $\times\ 12\arcsec$ ($PA = 120
\degr$) & [here] \\
\hline
$E(B-V)$ [mag]       & 0.062 & [2] \\
$V$ [mag]            & $18.07 \pm 0.10$ & \\
$V-I$ [mag]          & $0.34 \pm 0.10$ &  \\
$r_{\rm e}$ [arcsec] & $3.62 \pm 0.38$ & \\
SB$_{\rm 0,V}$ [mag\,arcsec$^{-2}$] & $22.0 \pm 0.2$ & \\
GALEX FUV [mag]      & $19.18 \pm 0.26$ & \\
GALEX NUV [mag]      & --- & \\
optical galaxy size  & $\sim$11 arcsec (370 pc) & \\ 
optical axis ratio   & $0.43 \pm 0.05$ & \\    
$PA$ [deg]           & $115 \pm 5$  & \\
\hline
\multicolumn{3}{l}{distance-dependent properties:} \\
\hline
$m - M$  [mag]         & $29.2 \pm 0.2$ & \\
$D$ [Mpc]              & $6.8 \pm 0.7$ & \\
$M_{\rm V}$ [mag]      & --11.27 & \\
$R_{\rm e}$ [pc]       & 118 & \\
log($M_{\rm *}$/\Msun) & 6.10 & \\
log(\MHI/\Msun)        & 7.08 & \\
log(SFR) [\MMsun\,yr$^{-1}$] & --3.04 & \\
log(sSFR) [yr$^{-1}$]  & --9.14 & \\
\hline 
\end{tabular}
{\flushleft {\small [1] \citet{Koribalski2018}, [2] \citet{Schlafly2011}. -- The GALEX FUV magnitude quoted here is the average of two estimates from the GALEX catalog.  
 }}
\label{tab:galaxy-prop}
\end{table}

\section{Environment}

We searched for neighboring galaxies in the Local volume ($D < 10$ Mpc) within a projected radius of 16.8 deg around Peekaboo corresponding to the linear projected separation of 2.0 Mpc at the distance 6.8 Mpc of the galaxy. We found the eight galaxies listed in Table~\ref{tab:ngc3621-group-tab1} \& Table~\ref{tab:ngc3621-group-tab2}. The columns of Table~\ref{tab:ngc3621-group-tab1} contain: (1)~galaxy name, (2)~equatorial coordinates, (3)~supergalactic coordinates, (4)~angular projected separation from Peekaboo, (5)~distance in Mpc and its uncertainty from EDD, measured via TRGB (only one galaxy, ESO\,379-G024, has an indirect distance estimate from Numerical Action Method = NAM, \citet{Shaya2017}, (6,7)~heliocentric radial velocity and radial velocity with respect to the Local Group centroid, (8)~peculiar velocity of the galaxy $v_{\rm pec} = v_{\rm LG} - H_0~D$ relative to the cosmic expansion with the Hubble parameter $H_0 = 73$\kms\,Mpc$^{-1}$, (9) apparent $B$- magnitude of the galaxy, taken from the UNGC collection, and (10) morphological type. The galaxies are ranked by their angular separation from Peekaboo, except two foreground galaxies: ESO\,379-G024 and ESO\,321-G014, belonging to the western outskirts of the nearby Cen~A group. The configuration of the remaining seven galaxies looks like a diffuse cloud with projected extensions of $1.8 \times 1.0$ Mpc and 1.6 Mpc extension in depth.  

\begin{table*}
\centering
\caption{Neighbourhood of the galaxy HIPASS~J1131--31 within a projected separation of 2.0 Mpc. }
\begin{tabular}{lccccccccc}
\hline
Galaxy & RA, DEC (J2000) & SGL, SGB & $r_{\rm p}$ & $D$ & $v_{\rm hel}$ & $v_{\rm LG}$ & $v_{\rm pec}$ & $B$ & Type \\
 & [hms, dms] & [deg, deg] & [deg] & [Mpc] & [\kkms] & [\kkms] & [\kkms] & [mag] &  \\
\hline
HIPASS J1131--31 & 11:31:34.6, --31:40:28 & 144.48, --25.71 & 0 &   $6.77 \pm 0.70$ & 716 & 433 & --71 & 18.2 & Irr  \\
HIPASS J1132--32 & 11:33:11.0, --32:57:43 & 145.93, --25.44 & 1.34 & $5.47 \pm 0.15$ & 699 & 414 & 5 & 16.3 & Irr  \\
NGC~3621 & 11:18:16.1, --32:48:42 & 145.65, --28.57 & 3.04 & $6.82 \pm 0.29$ & 730 & 440 & --70 & 10.2 & Scd  \\
ESO\,379-G007 & 11:54:43.0, --33:33:29 & 146.92, --21.00 & 5.20 & $5.33 \pm 0.17$ & 641 & 364 & --34 & 16.6 & Irr  \\
ESO\,320-G014 & 11:37:53.4, --39:13:14 & 152.86, --24.68 & 7.66 & $5.87 \pm 0.11$ & 654 & 362 & --77 & 15.9 & Irr  \\
ESO\,376-G016 & 10:43:27.1, --37:02:33 & 151.06, --35.48 & 10.92 & $6.75 \pm 0.33$ & 670 & 366 & --138  & 15.5 & Irr  \\
ESO\,318-G013 & 10:47:41.9, --38:51:15 & 153.12, --34.42 & 11.48 & $6.77 \pm 0.12$ & 718 & 413 & --93 & 15.0 & Sd  \\
\hline
ESO\,379-G024 & 12:04:56.7, --35:44:35 & 149.46, --19.12 & 8.05 & 3.4: (NAM) & 632 & 356 & -- & 16.6 & Irr  \\
ESO\,321-G014 & 12:13:49.6, --38:13:53 & 152.26, --17.62 & 10.84 & $3.24 \pm 0.16$ & 610 & 334 & 92 & 15.2 & IBm  \\
\hline
\end{tabular}
\label{tab:ngc3621-group-tab1}
\end{table*}

\begin{table*}
\centering
\caption{\HI\ properties of the galaxies listed in Table~3. --- The listed \MHI\  values are from LVHIS \citep{Koribalski2018}, re-calculated for the new distances.}
\begin{tabular}{lccccccc}
\hline
& & \multicolumn{1}{c}{LVHIS} & \multicolumn{2}{c}{HIPASS} & log  & log & log \\
Galaxy & HIPASS & \FHI  & \vhel & \wfi  & \MHI  & $L_B$ & SFR \\
~name & name & [Jy\kms] & [\kkms] & [\kkms] & [\MMsun] & [$L_{\odot}$] & [\MMsun yr$^{-1}$] \\
\hline
PGC~5060432   & HIPASS J1131--31 & 1.1 &  717 & 29  & 7.08 & 6.68 & --3.04 \\
PGC~0683190   & HIPASS J1132--32 & 1.4 & 699 & 59  & 7.01 & 7.27 & <--4.72  \\ 
NGC~3621      & HIPASS J1118--32 & 857  & 730 & 271 & 9.99  & 10.13 & 0.14 \\ 
ESO\,379-G007 & HIPASS J1154--33 & 4.8 & 641 & 25 & 7.52 & 7.14 & --3.62  \\
ESO\,320-G014 & HIPASS J1137--39 & 2.0 & 654 & 40 & 7.23 & 7.64 & --2.70  \\
ESO\,376-G016 & HIPASS J1043--37 & 10.3  & 668 & 33 & 8.06 & 7.74 & <--4.55 \\
ESO\,318-G013 & HIPASS J1047--38 & 8.6 & 711 & 42 & 7.98  & 8.00 & --2.05 \\
\hline
ESO\,379-G024 & HIPASS J1204--35 & 2.6 & 631 & 39  & 6.87 & 6.76 & <--5.08  \\
ESO\,321-G014 & HIPASS J1214--38 & 5.1  & 610 & 30 & 7.12 & 7.30 & --2.74  \\
\hline
\end{tabular}
\label{tab:ngc3621-group-tab2}
\end{table*}

Table~\ref{tab:ngc3621-group-tab2} lists the \HI\ properties of the same galaxies as in Table~\ref{tab:ngc3621-group-tab1}. Its columns contain: (1,2) galaxy name and its HIPASS name, (3) \HI\ flux  from LVHIS data, (4,5) heliocentric radial velocity and \HI\ line width at $50\%$ level of the maximum, (6) \HI\ mass, re-calculated for the new distances, (7) B-band luminosity given in UNGC, (8) integral star-formation rate, derived from the GALEX FUV magnitudes \citep{GildePaz2007}.

Due to the large projected separation of 370 kpc, Peekaboo cannot confidently be considered a true satellite of NGC~3621, although their radial distances are almost the same within uncertainties in $D_{TRGB}$.
The other five dwarf galaxies are also located far beyond the virial radius of NGC~3621, which is about 190 kpc. At the same time, the probability of an accidental grouping of these seven galaxies is extremely small. The radial velocity dispersion for the entities around NGC~3621 is only 56\kms. Formally applying the virial theorem, we obtain for this association a virial mass log$(M_{vir}/M_{\odot}) = 12.46$ and the ratio of the virial mass-to-B-band luminosity $M_{vir}/L_B = 210$. This loose non-virialized cloud has a crossing time of 14 Gyr, and resembles associations of dwarf galaxies like the nearby NGC~3109 group in Antlia \citep{Tully2006}.

\begin{figure} 
\centering
 \includegraphics[width=8cm]{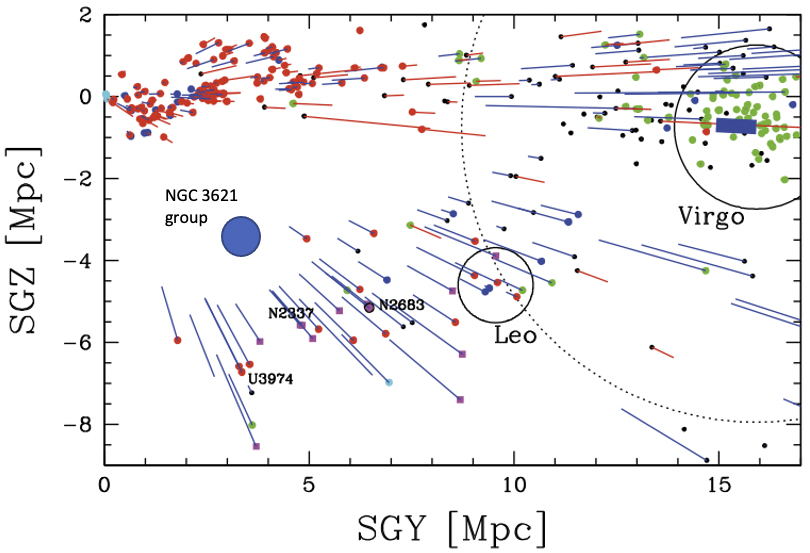}
\caption{This is the Figure~5 from \citet{Karachentsev2015} with the NGC~3621 group added at SGY = 3.5 Mpc and SGZ = --3.4 Mpc as the solid blue circle. }
\label{fig:K2015}
\end{figure}

The NGC~3621 association resides in a zone of very low spatial density, in a gap between the Local Sheet and the Dorado -- Leo Spur -- Antlia mini-wall \citep{Tully1988}. According to \citet{Karachentsev2015} and \citet{Anand2019}, these two nearly parallel planes, the Local Sheet and the Leo Spur, approach each other with a velocity of $\sim$260\kms, which is a reflection of the motion of the Local Sheet away from the Local Void. A projection of this region in supergalactic coordinates SGY, SGZ is shown in Figure~\ref{fig:K2015}. The Local Sheet extends from the Local Group towards the Virgo cluster at SGZ $\sim 0$. Negative peculiar velocity vectors in the Local Sheet rest frame are blue and directed toward us, while positive peculiar velocity vectors are red and directed away. A dotted circle indicates the Virgo cluster infall region. The NGC~3621 group at SGY = 3.5 Mpc and SGZ = $-3.4$ Mpc is located between the Local Sheet and the Leo Spur. The average peculiar velocity of galaxy NGC~3621 and its suite is $-70\pm17$\kms, demonstrating a moderate motion of the group toward the Local Sheet.

The small population of the NGC~3621 association is represented exclusively by late-type galaxies, as is typical for low-density volumes. Not a single dSph galaxy has been found in this region. The group galaxies have different specific star-formation rates, and the Peekaboo dwarf is the most active and most gas-rich object among them. We did not notice any correlation of the SFR or gas abundance with the projected separations of dwarf galaxies from NGC~3621.
 \\

\section{Is Peekaboo a young galaxy?}

 The compact dwarf galaxy Peekaboo with a predominantly young stellar population and very low metallicity is not unique in the Local Volume. \citet{Hirschauer2016}, \citet{McQuinn2020}, and \citet{Aver2022} investigated the properties of another compact dwarf galaxy AGC~198691 (= Leoncino), discovered in the ALFALFA blind \HI\ survey \citep{Haynes2011} and taken to lie at $D=12.0$~Mpc. In general appearance and stellar population, both galaxies look like twins. \citet{Sargent1970} named such objects "Isolated Extragalactic \HII\ regions". Their prototype is the famous dwarf system I~Zw~18 (= Mrk~116) with extremely low metallicity. \citet{Izotov2004} considered it as an example of a newly formed galaxy. With an angular diameter of $\sim$9 arcsec and a distance of $D = 18.2$ Mpc \citep{Aloisi2007}, I~Zw~18 has a linear diameter about 0.80 kpc.
Peekaboo, along with Leoncino and I\,Zw\,18, are XMP galaxies with HST imaging observations that bring into question whether they contain any ancient population of significance. In each case, the RGB is very insubstantial compared with the evident young populations.  \citet{Aloisi2007} point out that an identifiable RGB in I\,Zw\,18 argues for ages for these stars of at least 1~Gyr and probably $>2$~Gyr.  The same case can be made for the RGB stars in Leoncino and Peekaboo.  However this begs the question of whether there are stars in these galaxies that are ancient (ie, $>10$~Gyr).  
In Figure~\ref{fig:cmd}, isochrones for 3~Gyr and 10~Gyr are plotted that are indistinguishable as descriptions of the putative red giant branch.
In all other galaxies with sufficiently observed CMD (other than the likely tidally induced clumps around M81) prominent RGB are seen and provide probable evidence for ancient populations.  The situation with Peekaboo is decidedly ambiguous.  Arguably, star formation in this isolated XMP galaxy began long after the epoch of reionization. Peekaboo lies at half the distance of Leoncino and a third the distance of I\,Zw\,18 so moderate follow up observations with space telescopes should reach age sensitive populations.

\section{Concluding remarks}
We report on the optical and \HI\ properties of the nearby dwarf galaxy HIPASS~J1131--31 = Peekaboo. Images obtained with ACS/HST reveal a compact blue galaxy resolved into stars that was previously hidden in the halo of a bright foreground star. We estimate the galaxy distance to be $6.8 \pm 0.7$ Mpc. The galaxy has an absolute magnitude of  $M_V = - 11.27$, a half-light radius of $R_e = 120$ pc, and the central surface brightness of $SB_{0,V} = 22.0$. Peekaboo is a gas-rich dwarf galaxy with an \HI\ mass of  $\rm log(M_{\rm HI}/M_{\odot})= 7.08$ and a stellar mass of  $\rm log(M_*/M_{\odot}) = 6.10$. The ratio of its \HI-to-optical dimensions is about 2. From its detection in the GALEX FUV band, the galaxy is determined to have a specific star-formation rate of $\rm log(sSFR) = -9.14~yr^{-1}$.

Peekaboo is a likely member of the scattered association of late-type dwarf galaxies around the Scd galaxy NGC~3621 having dimensions of (1.0 -- 1.8) Mpc. This diffuse group is located in a very low density region between the Local Sheet and Dorado -- Leo Spur -- Antlia mini-wall, and moves toward the Local Sheet with a peculiar velocity of $-70 \pm 17$\kms.

Our optical spectral observations found Peekaboo to be one of the most metal-poor galaxies known with an oxygen abundance of 12+log(O/H) = $6.99\pm0.16$ {\bf (via the direct method) and 6.87$\pm$0.07 (via the strong lines method), i.e.} about $2\%$ of solar.  In its general shape, stellar population, and star formation rate, the compact Peekaboo dwarf galaxy looks like a twin of another star-forming dwarf galaxy AGC~198691 = Leoncino, investigated by \citet{Hirschauer2016} and \citet{McQuinn2020}. Both the dwarfs are extremely metal-poor objects with similarities to the better known I\,Zw\,18.  In all these cases, the RGB are scantily populated compared with the numbers of bluer young stars at similar magnitudes.  Plausibly, the small number of RGB stars that are seen are only a few Gyr old and these systems are devoid of ancient stellar populations. 


\vspace{-0.3cm}

\section*{Acknowledgements}

We thank Alec Hirschauer for useful critical comments.
This work is based on observations made with the NASA/ESA Hubble Space Telescope and with the Australia Telescope Compact Array.
The Australia Telescope Compact Array is part of the Australia Telescope National Facility which is funded by the Australian Government for operation as a National Facility managed by CSIRO. We acknowledge the Gomeroi people as the traditional owners of the Observatory site. Support for program SNAP-15992 (PI R.B.Tully) was provided by NASA through a grant from the Space Telescope Science Institute, which is operated by the Associations of Universities for Research in Astronomy, Incorporated, under NASA contract NASb5-26555. I.D.K. \& L.N.M. are supported in part by the Russian Federation grant 075-15-2022-233 (13.MNPMU.21.0003).
This work is also based on spectral observations obtained with the
Southern African Large Telescope (SALT), program \mbox{2021-1-MLT-001} (PI: Kniazev).
AYK acknowledges support from the National Research Foundation (NRF) of South  Africa.

\vspace{-0.3cm}

\section*{Data availability} 

The HST and ATCA data used in this article are available in the STScI archive and the AT Online Archive (ATOA\footnote{https://atoa.atnf.csiro.au/}), respectively. Additional data processing and analysis was conducted using the {\sc miriad} software\footnote{https://www.atnf.csiro.au/computing/software/miriad/} and the Karma visualisation\footnote{https://www.atnf.csiro.au/computing/software/karma/} packages. GALEX images were obtained through SkyView\footnote{https://skyview.gsfc.nasa.gov/current/cgi/query.pl}.










\section{Appendix: Other compact gas-rich dwarf galaxies in the Local Volume}

\begin{table*}
\centering
\caption{Properties of isolated compact gas-rich dwarf galaxies in the Local Volume}
\begin{tabular} {lccccc}
\hline
Name              &  HIDEEP~J1337--33 & Peekaboo & AGC749315 & AGC198507 & Leoncino  \\
PGC               & 677373 &       5060432 &         50596982    &     4194405 &       5807231  \\
\hline
R.A. [hms]      &  13 37 00.6 &    11 31 34.6 &      10 29 06.4 &     09 15 25.8 &   09 43 32.4  \\
Dec. [dms]      &   --33 21 47 &     --31 40 28 &        +26 54 38 &      +25 25 10 &    +33 26 58  \\
\vlg\ [\kkms] &      +371   &        +433  &            +564    &        +411 &         +464  \\   
\wfi\ [\kkms] &       20    &         29  &               31    &          37 &           33  \\
\FHI\ [Jy\,km\,s$^{-1}$] &      1.05  &         1.13 &             0.34    &        0.68 &   0.53 \\
$a_{\rm Ho}$ [arcsec] &    24       &      11   &              7        &      11   &          8  \\
$B$-band [mag] &    17.3     &      18.2 &             19.1  &          18.6     &     19.8   \\
FUV [mag] &    19.5     &      19.2     &         20.0  &          20.0     &     20.1  \\
$D$ [Mpc] &    4.55     &  6.90         &     9.46      &      10.96    &     12.0  \\
$log(M_{\rm HI})$ [\MMsun] &    6.71  &         7.10  &            6.86    &        7.28  &      7.26 \\
$A_{\rm Ho}$ [kpc]     &    0.53     &      0.38     &         0.32  &          0.57     &     0.47  \\
$M_{\rm B}$ [mag]        &    --11.2    &       --11.3   &          --10.9    &       --11.7   &      --10.6  \\
$log(L_B) [L_{\odot}]$  &       6.66  &          6.70     &         6.54  &          6.86     &     6.42  \\
$M_{HI}/L_B [M_{\odot}/L_{\odot}]$ & 1.1 & 2.5 & 2.1 & 2.6 & 6.9 \\ 
log(SFR) [$M_* yr^{-1}$] &   --3.52    &       --3.02   &          --3.17    &       --3.02  &       --3.06 \\

\hline
\multicolumn{6}{l}{Note. Sources of the distances: HIDEEP J1337--33 \citep{Grossi2007}, HIPASS~J1131--31 (present paper)}\\
\multicolumn{6}{l}{AGC~749315 \citep{Shaya2017}, AGC~198507 \citep{McQuinn2021}, Leoncino \citep{McQuinn2021}.}
\end{tabular}
\end{table*}

We take note of five dwarf galaxies in the Local Volume with $D < 12$~Mpc with linear Holmberg diameters $A_{\rm Ho}$ less than $\sim$0.5 kpc that, being rich in gas, were all found in the blind \HI\ surveys HIPASS and ALFALFA. The main parameters of these "intergalactic emission granules" are given in Table~5.
Here, we do not consider probable tidal galaxies: Clump~I, Clump~II, KDG61em with linear dimensions of $A_{\rm Ho}\sim(0.1-0.2)$~kpc and absolute magnitudes $M_B \simeq(-8,-9)$ mag, which occur in the \HI-shell of the interacting triplet M~81, M~82, NGC~3077 \citep{Karachentsev2011}, as well as "ragged" quenched ultra-compact dwarf objects, like the dE type satellite of the Sombrero galaxy, SucD1, with $A_{\rm Ho}\sim0.3$ kpc  \citep{Hau2009}, which are in close proximity to their massive host galaxies.

The galaxies in Table~5 are ranked by their distance from the Sun, and sources of distance estimates are given in the notes to the table. All distances, except AGC~749315, were measured by the TRGB method based on the images obtained with the HST. The distance of the dwarf galaxy AGC~749315 is estimated from its radial velocity, taking into account the local field of peculiar velocities in the manner described in the EDD. The table rows contain the following columns: (1,2) name of the galaxy and its number in HyperLEDA \citep{Makarov2014}; (3,4) equatorial coordinates: (5) radial velocity relative to the Local Group centroid; (6,7) \HI\ line width and the integral flux in this line; (8) angular Holmberg diameter in arcsec;
(9) apparent magnitude in the $B$-band; (10) apparent magnitude in the FUV-band of the GALEX survey \citep{Martin2005}; (11) distance of the galaxy; (12) logarithm of hydrogen mass; (13) linear diameter; (14) absolute $B$-magnitude; (15) logarithm of the $B$-band luminosity; (16) hydrogen mass-to-$B$-band luminosity ratio; (17) integral star formation rate determined from the FUV- flux. The parameters used and their sources are contained in the SAO LV database\footnote{SAO LV: sao.ru/lv/lvgdb}. The dwarf galaxies in the presented sample have a number of common properties.

i. The linear optical diameters of these blue compact objects are limited to the interval of [0.32 -- 0.57] kpc, which corresponds to the typical linear size of \HII\ regions inside spiral galaxies.

ii. The integral absolute magnitudes of the galaxies lie in the range of $\pm0.6$ mag relative to the median value $M_{\rm B}=  -11.2$ mag. Such galaxies are quite distinguishable up to the far edge of the LV, but their relative number in the LV population is only $\sim5/1200 = 0.4\%$.

iii. The considered dwarf galaxies are objects rich in gas. Their median ratio $ M_{\rm HI}/L_B$ equals to
2.5. Taking into account the presence of helium, about 90\% of the baryons in these galaxies are in the gaseous component. 

iv. Except for HIDEEP~J1337--33, which is a distant satellite of NGC~5236, the integral star formation rate of the remaining four dwarf galaxies lies in a surprisingly narrow range of [--3.02, --3.17] $M_{\odot}\,yr^{-1}$. These galaxies are able to reproduce their stellar mass during (1 - 3) Gyrs. Therefore, the objects under consideration can be called "youngish" galaxies.

As already noted, blue gas-rich compact dwarf galaxies with active star formation, like famous I~Zw~18, occur exclusively in regions of low spatial density. This circumstance requires explanation within the framework of the standard cosmological model.

\bsp	
\label{lastpage}
\end{document}